\documentclass{PoS}
\usepackage{amsmath,amssymb}

\title{Light quark correlators in a mixed-action setup}
\ShortTitle{Light quark correlators in a mixed-action setup}

\author{Fabio Bernardoni$^a$, Nicolas Garron \footnote{N.G. acknowledges supported from STFC grant ST/G000522/1 and EU grant 238353
(STRONGnet).}  $^b$, Pilar Hern\'andez$^c$,
  \speaker{Silvia Necco}$^d$, Carlos Pena$^e$\\

$^a$ NIC, DESY, Platanenallee 6, D-15738 Zeuthen, Germany\\

$^b$ SUPA, School of Physics, The University of Edinburgh, Edinburgh EH9 3JZ, UK\\

$^c$ Instituto de F\'isica Corpuscular, CSIC-Universitat de Val\`encia, Apartado de Correos 22085, E-46071 Valencia, Spain\\

$^d$ CERN, Physics Departement, 1211 Geneva 23, Switzerland \\

$^e$ Dpto. de F\'isica T\'eorica and Instituto de F\'isica T\'eorica UAM/CSIC, Universidad Aut\'onoma de Madrid, Cantoblanco E-28049 Madrid, Spain

\vspace{0.5cm}
E-mail: \email{Fabio.Bernardoni@desy.de, ngarron@staffmail.ed.ac.uk,
  pilar.hernandez@ific.uv.es, Silvia.Necco@cern.ch, carlos.pena@uam.es}
}

\abstract{
We report our progress in simulating Neuberger valence fermions on $N_f=2$ Wilson $O(a)$-improved sea quarks.
We compute correlators with valence quark masses both in the $p$- and in the $\epsilon$-regime, and we match the 
results with the predictions of the Chiral Effective Theory in the mixed regime. This allows us to extract the 
Low Energy Couplings (LECs) of the $N_f=2$ theory and to test the validity of the approach.

\vspace{1.5cm}
\begin{flushright}
CERN-PH-TH/2011-229\\
IFT-UAM/CSIC-11-68\\
Edinburgh 2011/28\\
FTUAM-11-55\\
DESY 11-171\\
IFIC/11-54
\end{flushright}
         }

\FullConference{The XXIX International Symposium on Lattice Field Theory - Lattice 2011\\
July 10-16, 2011\\
Squaw Valley, Lake Tahoe, California}

\begin{document}
\section{Introduction}
Lattice QCD simulations with different discretizations in the valence and in the sea sector have been performed by several groups and with different setups in the last years \cite{Renner:2004ck,Allton:2006nu,Durr:2007ef,Aubin:2008wk,Li:2010pw,Cichy:2010ta,Beane:2011zm}. This is a viable approach to preserve exact chiral symmetry for the valence quarks with a ``moderate'' numerical cost, which could be particularly advantageous for the computation of weak matrix elements.

In our work we apply the Neuberger operator on $N_f=2$ non-perturbatively $O(a)$-improved Wilson fermions configurations generated within  the CLS common effort. \footnote{https://twiki.cern.ch/twiki/bin/view/CLS/WebHome} In a recent publication \cite{Bernardoni:2010nf} we have reported our results on the topological susceptibility and the low-lying eigenvalues of the massless Dirac operator, which we used to test the so-called mixed Chiral Effective Theory approach and finally to extract the chiral condensate.
We now extend our analysis to two- and three-point functions.

\section{Left current correlator in Chiral Effective Theory}
We consider partially quenched QCD with $N_s$ sea flavors with mass $m_s$ and $N_v$ valence flavors with mass $m_v$, and define the two-point correlators of the left-handed current
\begin{equation}\label{eq:corr}
{\rm Tr}[T^aT^b]C(x_0)\equiv \int d^3x \langle J^a_0(x)J^b_0(0) \rangle,\;\;\;\;\;J^a_\mu\equiv \bar{\psi}T^a\gamma_\mu P_-\psi,
\end{equation}
where $T^a$ is a traceless generator of $SU(N_v)$ and $P_-=(1-\gamma_5)/2$ is the left projector. 
Low energy properties can be described by the corresponding $SU(N_s+N_v)$ Partially Quenched Chiral Perturbation Theory (PQChPT); in particular, in a finite volume $V=L^3 T$ we consider the following two situations \cite{Bernardoni:2007hi}:
\begin{itemize}
\item $m_{v/s}\Sigma V\gg 1$: all quarks are in the $p$-regime. In this case the correlator $C(x_0)$ at NLO in PQChPT is given by
\begin{equation}\label{eq:corr1}
\mathcal{C}(x_0)=\frac{F^2_{\rm{eff}}}{4}\frac{M_{vv;\rm{eff}}\cosh\left[M_{vv;\rm{eff}}(T/2-x_0)  \right]   }{\sinh\left[M_{vv;\rm{eff}}T/2\right]}-\frac{N_s}{2}\frac{dG^1(0;M^2_{vs})}{dT},
\end{equation}
with
\begin{eqnarray}
F^2_{\rm eff}  = F^2(1+\Delta_F)&=&F^2\left\{1-\frac{N_s}{F^2}G^1(0;M^2_{vs})+\frac{8}{F^2}(N_sM^2_{ss}L_4+M^2_{vv}L_5)\right\},\label{eq:feffp}\\
M^2_{vv;\rm{eff}} = M^2_{vv}(1+\Delta_M)&=&M^2_{vv}\Bigg\{1+\frac{1}{N_sF^2}\left[G^1(0;M^2_{vv})+(M^2_{ss}-M^2_{vv})G^2(0;M^2_{vv})   \right]\label{eq:meffp}  \nonumber\\
& - & \frac{8}{F^2}\left[N_sM^2_{ss}(L_4-2L_6)+M^2_{vv}(L_5-2L_8)\right]\Bigg\},
\end{eqnarray}
with the LO pseudoscalar masses $M_{ij}=(m_i+m_j)B$ and the pseudoscalar propagators in finite volume
\begin{equation}
G^r(x;M^2)=\frac{1}{V}\sum_p\frac{e^{ipx}}{(p^2+M^2)^r},\;\;\;r\geq 1.
\end{equation}
\item $m_v \Sigma V\lesssim 1$, $m_s\Sigma V\gg 1$: valence quarks are in the $\epsilon$-regime, sea quarks are in the $p$-regime.
In this case the NLO left correlator at fixed topology is given by
\begin{equation}\label{eq:corr2}
\mathcal{C}^\nu(x_0)=\frac{F^2_{\epsilon;\rm{eff}}}{2T}\left\{1+\frac{2T^2}{F^2V}\mu_v\sigma_\nu(\mu_\nu)h_1\left(\frac{x_0}{T}\right)  \right\}-\frac{N_s}{2}\frac{dG^1(0;M^2_{ss}/2)}{dT},
\end{equation}
with $\mu_v=m_v\Sigma V$ and
\begin{eqnarray}
F^2_{\epsilon;\rm{eff}} & =& F^2\left(1+\Delta_F|_{M_{vv}=0}\right)= F^2\left\{1-\frac{N_s}{F^2}G^1(0;M^2_{ss}/2)+\frac{8}{F^2}(N_sM^2_{ss}L_4)\right\},\label{eq:feffe}\\
h_1(\tau)& = & \frac{1}{2}\left[\left(\tau-\frac{1}{2}\right)^2 -\frac{1}{12}  \right],\\
\sigma_\nu(x) & = & x\left[I_\nu(x)K_\nu(x)+I_{\nu+1}(x)K_{\nu-1}(x)\right]+\frac{\nu}{x},
\end{eqnarray}
where $I_\nu$, $K_\nu$ are modified Bessel functions.
\end{itemize}
A well-established strategy to extract the LO ($F,B$) and the NLO ($L_i$) Low Energy Couplings (LECs) is to match lattice results for the two-point correlators with the predictions of the ChPT. Depending on the chiral regime, the systematic uncertainty associated to the matching can vary considerably. Exploring different regimes in a finite box can then be useful to give reliable estimates of systematic errors associated to the LECs. 
\section{Lattice results}
\subsection{Setup and results for the Low Energy Couplings}
We have carried out our simulations on CLS lattices of size $V=24^3\times 48$, with $\beta=5.3$ corresponding to a lattice spacing $a= 0.0649(10) $ fm \cite{Marina}. We have considered three $p$-regime sea quark masses $\kappa=0.13620,0.13625,0.13635$, labeling the sets by D4,D5,D6 respectively. On these configurations we have built the Neuberger-Dirac operator, with valence quark masses both in the $p$- and in the $\epsilon$-regime. The simulations parameters are collected in Table \ref{tab:sim}, where we report also the effective sea pion masses in lattice units (see \cite{Bernardoni:2010nf} for further details). For valence quark masses
in the $p$-regime, we have collected $\sim 80 - 160$ measurements. For the $\epsilon$-regime, we have in total $\sim 70-100$ measurements of the left correlator in the topological sectors $|\nu|=0,1,2,3$.
\begin{table}
\begin{center}
\begin{tabular}{|l|l|l|l|l|}
\hline
\multicolumn{5}{|l|}{$\beta=5.3$, \hspace{0.5cm} $c_{SW}=   1.90952$,     \hspace{0.5cm} $V=24^3 \times 48$      } \\
\hline
lattice & $\kappa$ & $aM_{ss;{\rm eff}}$ & $am_v$ ($\epsilon$-reg.) & $am_v$ ($p$-reg.) \\
\hline 
D4 & 0.13620 & 0.1695(14) & 0.002 & 0.012,0.02,0.028 \\
D5 & 0.13625 & 0.1499(15) & 0.002 & 0.012,0.02 \\
D6 & 0.13635 & 0.1183(37) & 0.002 & 0.006,0.015,0.025 \\
\hline
\end{tabular}
\end{center}
\caption{Simulation parameters}\label{tab:sim}
\end{table}

\begin{figure}
\includegraphics[width=7cm]{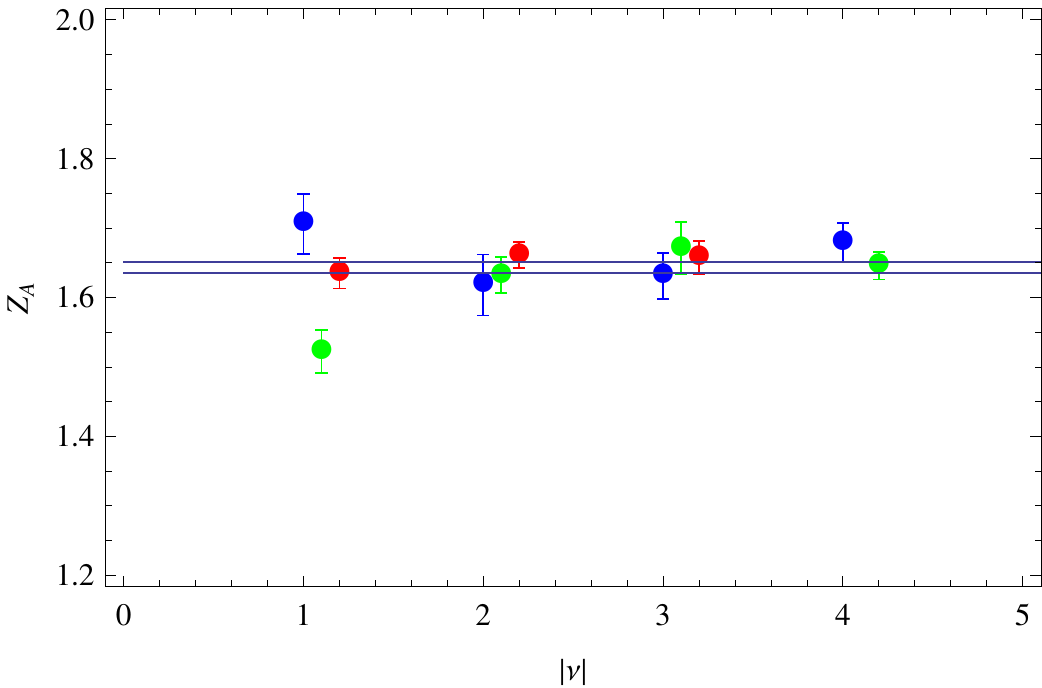}
\includegraphics[width=7cm]{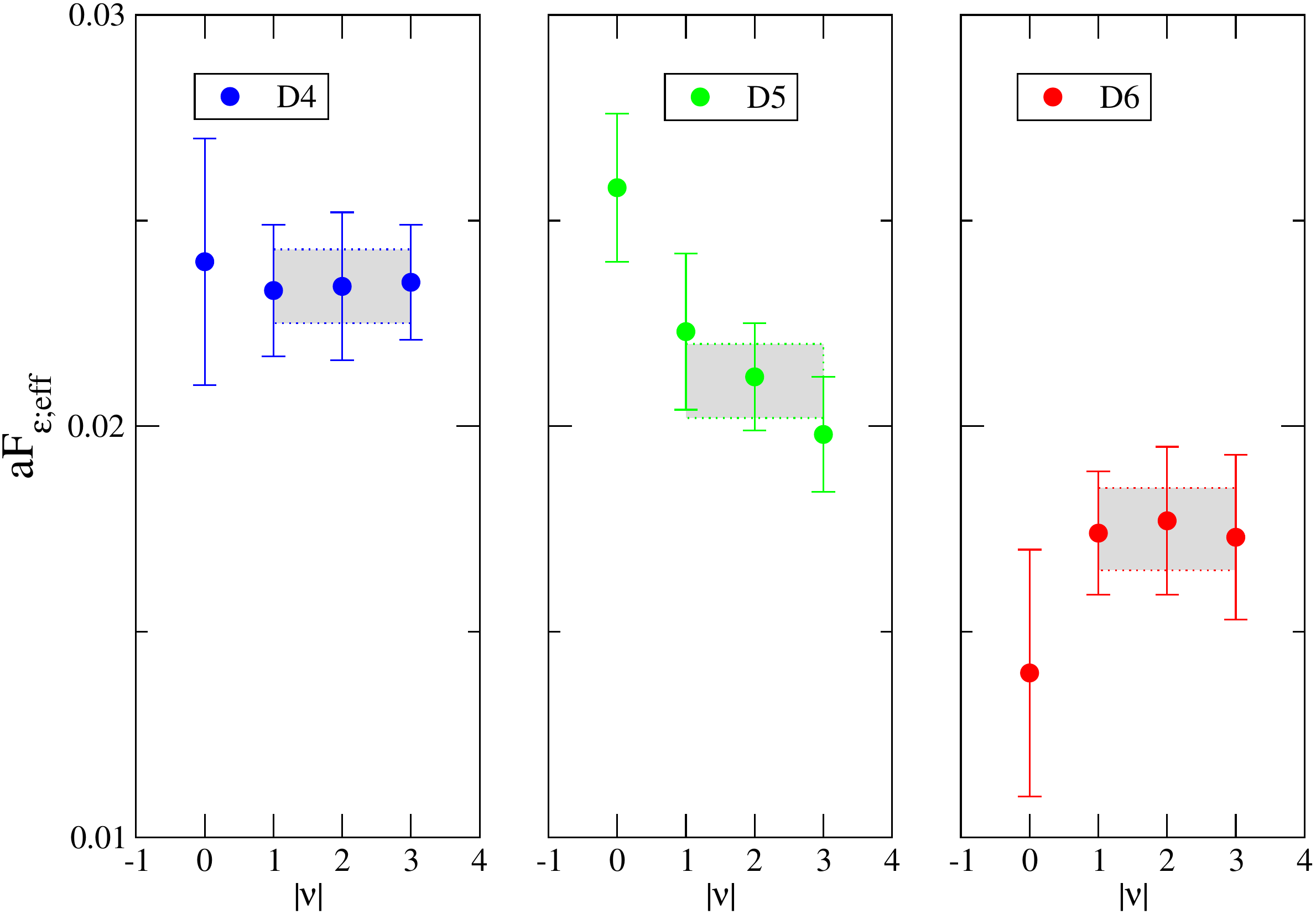}
\caption{\emph{Left}: the renormalization constant of the local axial current extracted through the chiral Ward identity at fixed topological charge saturated with the zero modes. The blue,green and red points refer to the lattices D4,D5,D6 respectively. 
\emph{Right}: $aF_{\epsilon;\rm{eff}}$ extracted from the left correlator in the mixed  $p$/$\epsilon$-regime as a function of the topological charge. The gray bands represent the average over $|\nu|=1,2,3$. 
}\label{fig:za}
\end{figure}

We have computed the two-point left current correlator defined in Eq.~(\ref{eq:corr}) adopting the low-mode averaging technique of \cite{Giusti:2004yp}. 
For the renormalization of the local axial current we used the chiral Ward identity at fixed topological charge saturated with the zero modes, following the strategy proposed in \cite{Giusti:2003iq}. In particular we considered
\begin{equation}
Z_A^{\rm mix}\lim_{m_v\rightarrow 0}(m_vV)\sum_{\vec{x}}\langle \partial_0P^a(x)J^a_0(0)\rangle_\nu =\lim_{m_v\rightarrow 0}(m_v^2V)\sum_{\vec{x}}\langle P^a(x)P^a(0)\rangle_\nu,
\end{equation}
where $P^a=\bar{\psi}T^a\gamma_5\psi$ is the pseudoscalar density.
The right-hand side is completely saturated with zero modes and can be directly computed without chiral extrapolation; on the left-hand side, the chiral limit has to be taken numerically. The renormalization factor $Z_A^{\rm mix}$ can then be extracted from ratios of correlations functions. For those correlators we have collected a total of $\sim 70-140$ measurements for the topological sectors $0 < |\nu| < 5$.
On the left side of Fig.~\ref{fig:za} we report the results for the different sea quark masses corresponding to D4,D5,D6 and for  $\nu=1,2,3,4$.
Apart from some fluctuations in the topological sector $|\nu|=1$, the data do not depend on $\nu$ and on the sea quark masses. From this study we obtained
\begin{equation}
Z_A^{\rm mix}=1.65(1).
\end{equation}
We then extracted $M_{vv;\rm{eff}}$, $F_{\rm eff}$, $F_{\epsilon;\rm{eff}}$ from the left correlators, according to Eqs.~(\ref{eq:corr1}, \ref{eq:corr2}); on the right side of Fig.~\ref{fig:za} we show for instance the (bare) pseudoscalar decay constant computed from the mixed $p$/$\epsilon$-regime correlators at $|\nu|=0,1,2,3$. As it should be, results do not depend significantly on the topological charge; $F_{\epsilon;\rm{eff}}$ has been determined by averaging over $|\nu|=1,2,3$ (gray band in the plot).  
 We then performed a global fit of pseudoscalar masses and decay constants using the NLO predictions of the PQChPT given in Eqs.~(\ref{eq:feffp},\ref{eq:meffp},\ref{eq:feffe}). From this fit we obtained the six parameters 
\begin{eqnarray} 
aF^R=0.0259(10)(8) &,& aB^R=1.52(11)(9),\nonumber\\
L_4^R = 0.00054(10)(4) &,&  L_5^R=0.00073(15)(2),\label{eq:fitres}\\
(L_4-2L_6)^R = 0.00033(5)(0) &,& (L_5-2L_8)^R = -0.00025(15)(5),\nonumber
\end{eqnarray}
where ``R'' means that the couplings are renormalized.
The constants $L_i^R$ are given at the scale $\mu=770$ MeV, while $B^R$ is given in the $\overline{\rm{MS}}$ scheme at $\mu=$ 2 GeV.
For the tree-level pseudoscalar masses we used $M_{ij}=(m_i^R+m_j^R)B^R$; for the sea Wilson quark masses we used the PCAC masses computed in \cite{DelDebbio:2007pz,Marina}, with renormalization constants $Z_A^W=0.78(1)$ \cite{DellaMorte:2008xb}, $Z_P^W(\overline{\rm MS},\; 2\; \rm{GeV})=0.543(8)$ \cite{DellaMorte:2005kg}; for the overlap valence quarks we have computed $Z_S^{\rm mix}(\overline{\rm MS},\; 2\; \rm{GeV})=1.88(15)$ like in \cite{Bernardoni:2010nf}. 
The first error is statistical, while the second one reflects the uncertainty on the PCAC quark masses.
\begin{figure}
\includegraphics[width=8cm]{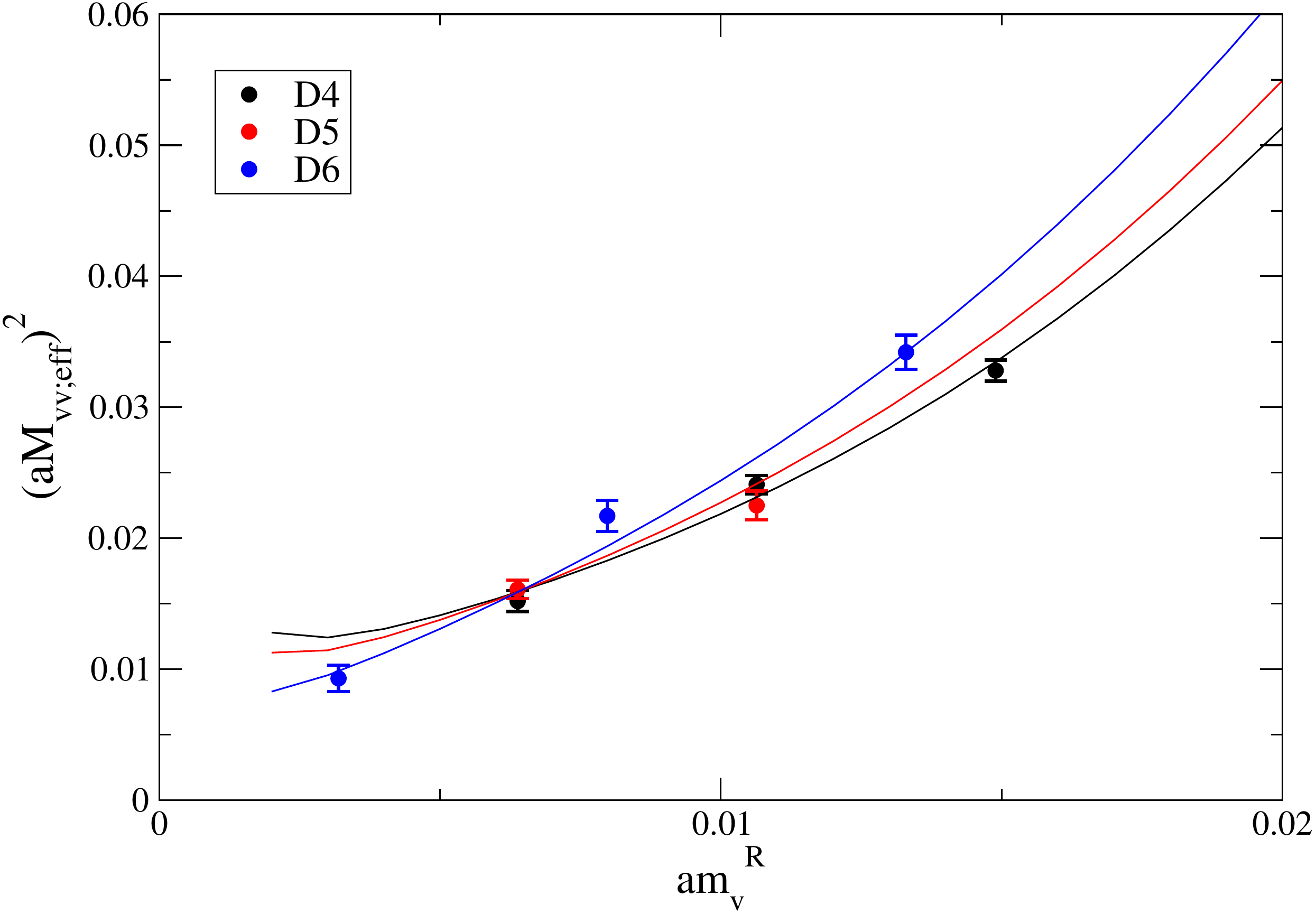}
\includegraphics[width=8cm]{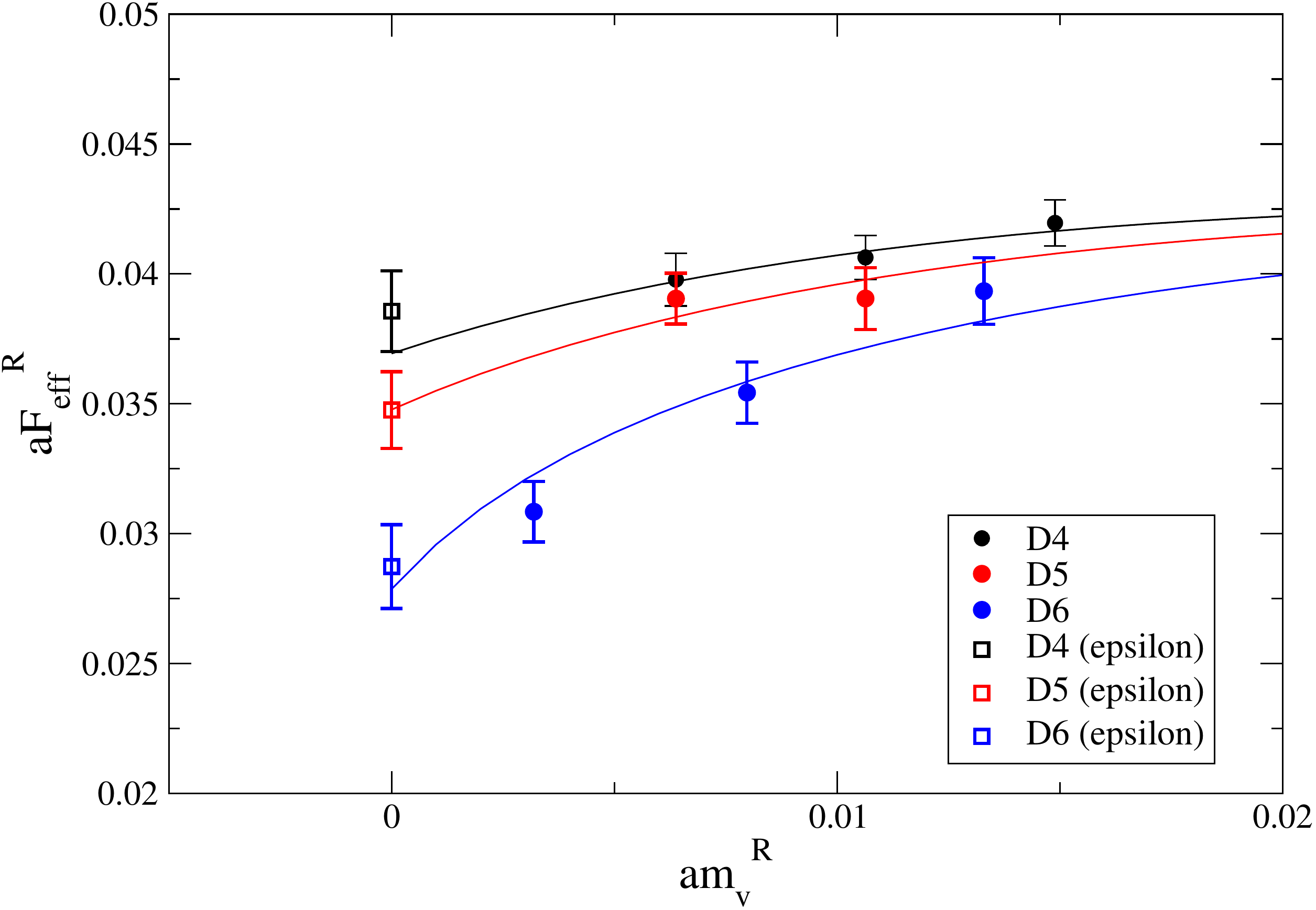}
\caption{Global chiral NLO fit for $M_{vv;\rm{eff}}$, $F_{\rm eff}$. The squares in the right plot correspond to the $\epsilon$-regime results, $F_{\epsilon;\rm{eff}}$.} \label{fig:fit}
\end{figure}

The global fit is shown in Fig.~\ref{fig:fit}. It is important to stress that having valence quark masses in the $\epsilon$-regime allows us to extract the pseudoscalar decay constant directly at $am_v=0$ and hence to have a good control over the chiral extrapolation in $m_v$. \\
Appropriate combinations of the $L_i$ yield the conventional SU(2) LECs $\bar{l}_i$; in particular we obtain 
\begin{equation}
\bar{l}_3=4.46(30)(14),\;\;\;\;\;\;\;\;\bar{l}_4=4.56(10)(4).
\end{equation}
The FLAG working group \cite{Colangelo:2010et} gives an estimate $\bar{l}_3=3.2(8)$ and quotes $\bar{l}_4=4.66(4)\binom{+4}{-33}$ \cite{Baron:2009wt} as best determination of $\bar{l}_4$. Our determinations are in agreement with those results within the uncertainties. By converting $aF^R$ and $aB^R$ in physical units we obtain $F^R=79(4)$ MeV and $\Sigma^R(\overline{\rm MS},\; 2\; \rm{GeV})=B^R(F^R)^2=(306(11)\;{\rm MeV})^3$, which also lie in the right ballpark. 
\subsection{Unitary point}
\begin{figure}
\begin{minipage}{7.5cm}
\includegraphics[width=7.5cm]{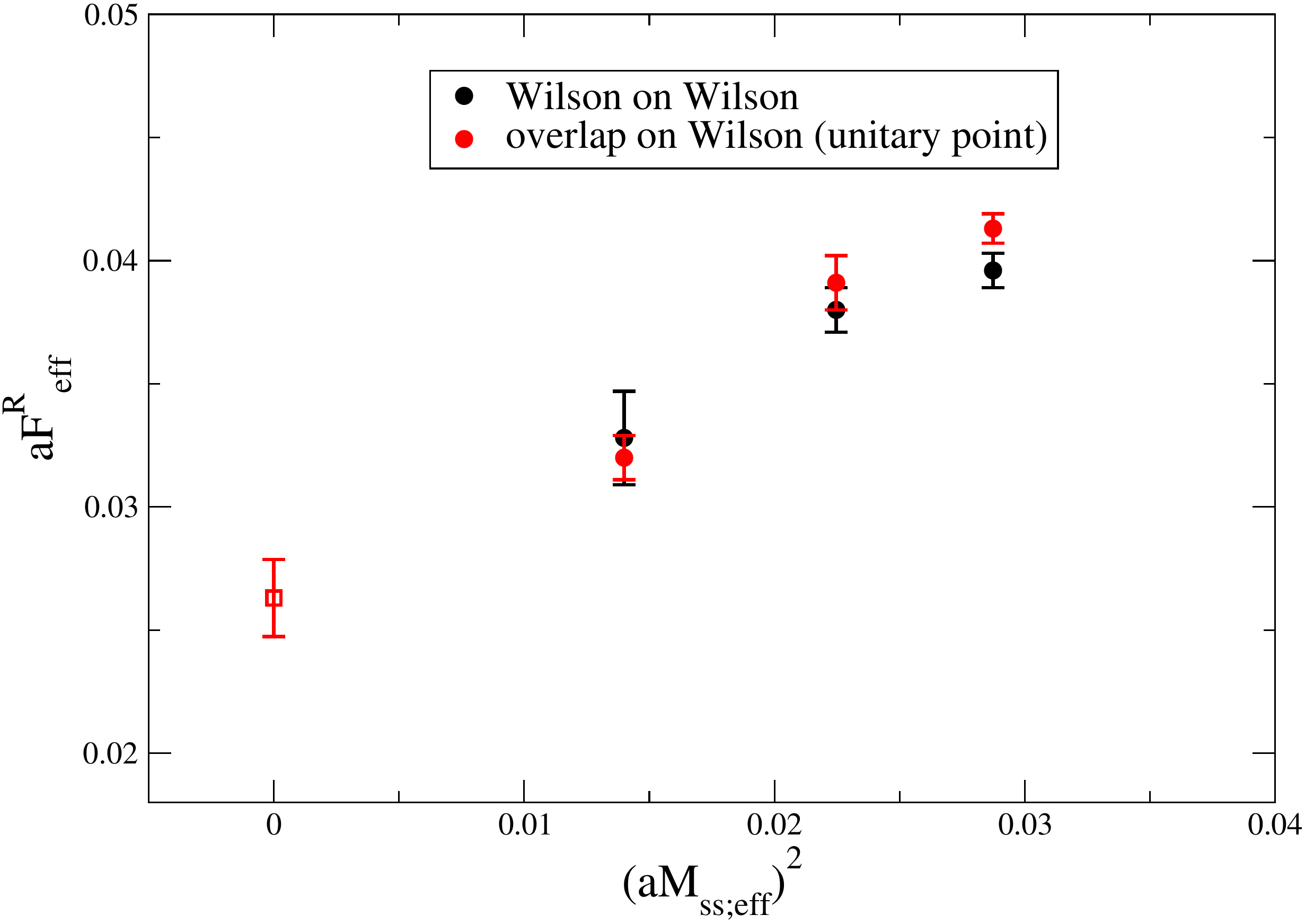}
\end{minipage}
\begin{minipage}{6cm}
\begin{tabular}{llll}
\hline
lat. & $am_{v;\rm{un}}$ & $F^R_{\rm eff}|_{m_v=m_{v;\rm{un}}}$  & $F^R_{s;\rm{eff}}$  \\
\hline
D4 & 0.0241(5)(4)& 0.0413(6)& 0.0396(7)\\
D5 & 0.0196(13)(5)& 0.0391(11)& 0.0380(9)\\
D6 & 0.0088(6)(6) & 0.0320(9) & 0.0328(19)\\
\hline
\end{tabular}
\end{minipage}
\caption{Effective pseudoscalar decay constant computed in the full Wilson theory (black points) and in the mixed action theory at the unitary point (red points) as a function of the pseudoscalar mass. The empty square represents the decay constant in the chiral limit extracted from the chiral fit, Eq.~(\protect\ref{eq:fitres}).} \label{fig:unit}
\end{figure}

A natural definition of the unitary point corresponds to the matching of the pseudoscalar masses
\begin{equation}
M^2_{vv;\rm{eff}}|_{m_v=m_{v;\rm{un}}}=M^2_{ss;\rm{eff}}.
\end{equation}
We have extracted $m_{v;\rm{un}}$ through a linear interpolation of $M^2_{vv;\rm{eff}}$. 
The pseudoscalar decay constants in the full Wilson theory ($F^R_{s;\rm{eff}}$) and in the mixed action theory at $m_{v;\rm{un}}$ ($F^R_{\rm eff}|_{m_v=m_{v;\rm{un}}}$) should then be equal up to discretization effects. 
In Fig.~\ref{fig:unit} we show $F^R_{\rm eff}|_{m_v=m_{v;\rm{un}}}$ and $F^R_{s;\rm{eff}}$ for the three lattices D4,D5,D6. The Wilson on Wilson data have been taken from \cite{DelDebbio:2007pz}. Within our statistical accuracy, we do not observe large cutoff effects close to the unitary point. 

Discretization effects in a mixed action setup can be investigated by means of Mixed Action PQChPT \cite{Bar:2002nr}.
If all mass scales are in the $p$-regime and one assumes a power counting $a\sim m_{s,v}\sim p^2$ (\emph{generically small masses}), it is easy to see that ``unitarity violations'' of $O(a^2)$ appearing in $M^2_{vs}$ and $(M^2_{ss}-M^2_{vv})$ affect our observables only at NNLO, and hence are expected to be small. Our numerical data seem to confirm this expectation, although a more detailed study is necessary in order to have a precise estimate of those effects. \footnote{A recent study with Neuberger valence fermions and Wilson Twisted Mass sea quarks \cite{Cichy:2010ta} indicates large cutoff effects close to the unitary point.}

\section{$K\rightarrow \pi\pi$ amplitudes and $\Delta I=1/2$ rule}
In \cite{Giusti:2004an}, a strategy to study the r\^ole of the charm quark in the $\Delta I=1/2$ rule has been proposed. 
The basic idea is to start from the GIM limit $m_u=m_d=m_s=m_c$, write down the $\Delta S=1$, CP-even effective weak Hamiltonian and match it with the effective Hamiltonian in the SU(4) Chiral Effective Theory. At leading order in ChPT, the ratio of amplitudes corresponding to $\Delta I=1/2$ and $\Delta I=3/2$ $K\rightarrow \pi\pi$ transitions can be expressed in terms of given LECs $g_1^{\pm}$. As usual, those couplings can be computed by matching suitable correlation functions computed on the lattice with their analytic expressions obtained in ChPT (see \cite{Giusti:2004an} for details).
Quenched results obtained with Neuberger fermions have been presented in \cite{Giusti:2006mh}; this strategy, for which exact chiral symmetry at finite lattice spacing is an essential ingredient, can be implemented also with our mixed action setup. Preliminary results for the bare ratios $R^{\pm}$ of three- and two-point functions defined in \cite{Giusti:2006mh} have been obtained so far for the lattice D4 (interpolated to the unitary point)
\begin{equation}
R^{+}=0.64(6),\;\;\; R^{-}=1.5(2), 
\end{equation}
of the same order as the quenched results for pseudoscalar masses in the same range. \\
The second part of the strategy consists in studying the amplitudes ratio as a function of $m_c > m_u = m_d = m_s$ and test if 
by increasing the charm quark the $\Delta I=1/2$ enhancement observed experimentally is reproduced.


\section{Conclusions and outlook}
Mixed action simulations with exact chiral symmetry in the valence sector provide a powerful tool to investigate QCD properties. 
Given the significant errors associated to chiral extrapolations, it is important to explore several
chiral regimes where the systematic effects are different, so that such errors can be properly quantified.
In particular, we have considered valence quarks both in the $p$- and in the $\epsilon$-regime and matched the lattice results with the PQChPT predictions at NLO. A global chiral fit of pseudoscalar masses and decay constants allowed us to determine $F$, $B$, $\bar{l}_3$, $\bar{l}_4$ with comparable accuracy with respect to other results present in the literature. By matching pseudoscalar masses in the valence and in the sea sector, we did not observed large cutoff effects close to the unitary point.
We stress that a precise determination of the LECs would require an improvement of the statistical precision and a careful monitoring of autocorrelations; furthermore, a larger set of volumes, lattice spacings and quark masses would guarantee a solid estimate of the systematic uncertainties associated to lattice artefacts, finite-volume effects and chiral extrapolation. 

As next step, we are implementing the strategy proposed in \cite{Giusti:2004an} to investigate the r\^ole of the charm quark in the $\Delta I=1/2$ rule. This calculation is more challenging and requires further computational efforts.


\begin{thebibliography}{10}

\bibitem{Renner:2004ck}
LHP Collaboration, D.B. Renner et~al.,
\newblock Nucl.Phys.Proc.Suppl. 140 (2005) 255.

\bibitem{Allton:2006nu}
C. Allton et~al.,
\newblock PoS LAT2006 (2006) 202.

\bibitem{Durr:2007ef}
S. D\"urr et~al.,
\newblock PoS LAT2007 (2007) 115.

\bibitem{Aubin:2008wk}
C. Aubin, J. Laiho and R.S. Van~de Water,
\newblock Phys.Rev. D77 (2008) 114501; J. Laiho and R.S. Van~de Water,
\newblock PoS LATTICE2010 (2010) 312.

\bibitem{Li:2010pw}
xQCD Collaboration, A. Li et~al.,
\newblock Phys.Rev. D82 (2010) 114501.

\bibitem{Cichy:2010ta}
K. Cichy, G. Herdoiza and K. Jansen,
\newblock Nucl.Phys. B847 (2011) 179.

\bibitem{Beane:2011zm}
S. Beane et~al.,
\newblock (2011), 1108.1380 [hep-lat].


\bibitem{Bernardoni:2010nf}
F. Bernardoni et~al.,
\newblock Phys.Rev. D83 (2011) 054503.

\bibitem{Bernardoni:2007hi}
F. Bernardoni and P. Hern\'andez,
\newblock JHEP 10 (2007) 033.

\bibitem{Marina}
M. Marinkovic et~al.,
\newblock PoS Lattice2011 (2011) 232.

\bibitem{Giusti:2004yp}
L. Giusti et~al.,
\newblock JHEP 0404 (2004) 013;
T.A. DeGrand and S. Schaefer,
\newblock Comput.Phys.Commun. 159 (2004) 185.

\bibitem{Giusti:2003iq}
L. Giusti et~al.,
\newblock JHEP 01 (2004) 003; P. Hern\'andez et~al.,
\newblock JHEP 0805 (2008) 043.

\bibitem{DelDebbio:2007pz}
L. Del~Debbio et~al.,
\newblock JHEP 02 (2007) 082.

\bibitem{DellaMorte:2008xb}
M. Della~Morte, R. Sommer and S. Takeda,
\newblock Phys.Lett. B672 (2009) 407.

\bibitem{DellaMorte:2005kg}
ALPHA Collaboration, M. Della~Morte et~al.,
\newblock Nucl.Phys. B729 (2005) 117.

\bibitem{Colangelo:2010et}
G. Colangelo et~al.,
\newblock Eur.Phys.J. C71 (2011) 1695.

\bibitem{Baron:2009wt}
ETM Collaboration, R. Baron et~al.,
\newblock JHEP 1008 (2010) 097.

\bibitem{Bar:2002nr}
O. B\"ar, G. Rupak and N. Shoresh,
\newblock Phys.Rev. D67 (2003) 114505; O. B\"ar, G. Rupak and N. Shoresh,
\newblock Phys.Rev. D70 (2004) 034508; M. Golterman, T. Izubuchi and Y. Shamir,
\newblock Phys.Rev. D71 (2005) 114508; O. B\"ar, M. Golterman and Y. Shamir,
\newblock Phys.Rev. D83 (2011) 054501.

\bibitem{Giusti:2004an}
L. Giusti et~al.,
\newblock JHEP 11 (2004) 016.

\bibitem{Giusti:2006mh}
L. Giusti et~al.,
\newblock Phys.Rev.Lett. 98 (2007) 082003.

\end{thebibliography}

\end{document}